\documentclass[preprint]{elsarticle}
\usepackage{amsmath,amssymb}
\usepackage{slashed}
\usepackage{graphicx}%
\usepackage{subfigure}
\usepackage{mathptmx}
\usepackage{bm}% bold math
\newcommand{\beq}{\begin{equation}}
\newcommand{\eeq}{\end{equation}}

\newcommand{\be}{\begin{eqnarray}}
\newcommand{\ee}{\end{eqnarray}}
\long\def\hidestart#1\hideend{}
\begin{document}

\title{On the critical exponent of $\eta/s$ and a new exponent-less measure 
of fluidity}

\author{Raktim Abir}
\ead{raktim.abir@saha.ac.in}

\author{Munshi G. Mustafa}
\ead{munshigolam.mustafa@saha.ac.in}

\address{Theory Division, Saha Institute of Nuclear Physics \\
 1/AF Bidhan Nagar, Kolkata 700064, India}

\begin{keyword}
Relativistic Fluid, Quark-Gluon Plasma, Heavy-Ion Collision, Viscosity, Thermal Conductivity
%\PACS 12.38.Mh, 47.75.+f, 52.35.Mw, 47.35.-i
\end{keyword} 
\date{}
\begin{abstract}
We discuss on the critical exponent of $\eta/s$ for a fluid, and propose 
a new exponent-less measure of fluidity based on a mode-mode coupling theory. 
This exhibits a remarkable universality for fluids obeying a liquid-gas phase 
transition both in hydrodynamic as well as in nonhydrodynamic region. We 
show that this result is independent of the choice of the fluid dynamics, 
{\em viz.}, relativistic or nonrelativistic. Quark-Gluon Plasma, being a hot 
relativistic and a nearly perfect fluid produced in relativistic heavy-ion 
collisions, is expected to obey the same universality constrained by both 
the viscous and the thermal flow modes in it. We also show that if 
the elliptic flow data in RHIC puts a constraint on $\eta/s$ then the new 
fluidity measure for Quark-Gluon Plasma in turn also restricts the other
transport coefficient, {\it viz.}, the thermal conductivity. 
\end{abstract}

\maketitle
%%%%%%%%%%%%%%%%%%%%%%%%%%%%%%%%%%%%%%%%%%%%%%%%%%%%%%%%%%
%\section{}\label{intro}
%%%%%%%%%%%%%%%%%%%%%%%%%%%%%%%%%%%%%%%%%%%%%%%%%%%%%%%%%%

%%%%%%%%%%%%%%%%%%%%%%%%%%%%%%%%%%%%%%%%%%%%%%%%%%%%%%%%%%%%%%%%%
%\section{}
%%%%%%%%%%%%%%%%%%%%%%%%%%%%%%%%%%%%%%%%%%%%%%%%%%%%%%%%%%%%%%
Recent results~\cite{white,ellip} from the Relativistic Heavy 
Ion Collider (RHIC) at BNL reveal surprising and intriguing dynamical 
properties of the Quark-Gluon Plasma (QGP). In non-central collisions, 
the anisotropy with respect to the reaction plane, {\it i.e.}, the 
elliptic flow coefficient $v_2$\cite{ellip}, can well be 
described up to transverse momenta of order $1.5$ GeV/$c$ by the nearly 
ideal hydrodynamics with small shear viscosity, $\eta$. This is much
smaller than that obtained in the perturbative quark-gluon plasma~\cite{pqcd}.
This suggests that the matter produced in the early phase of the RHIC
collisions is strongly interacting with nearly perfect fluidity represented 
by a very small ratio of shear viscosity to entropy density, $\eta/s$. The 
supporting arguments for it have come from various directions: {\it viz}, 
viscous hydrodynamics~\cite{heinz,asis} sets an upper bound, 
$\eta/s \leq (5/4\pi)$; the gauge-gravity dual theory~\cite{kss} 
conjectures at universal lower bound, $\eta/s\geq1/4\pi$; and the Heisenberg's 
uncertainty principle implies a lower bound~\cite{kss,schafer}, $\eta/s\sim 1$ 
(we use units with $\hbar =k_B=c=1$). The fluidity of QCD fluid has 
also been computed from numerical simulation in lattice~\cite{lattice}, 
non-perturbative 
models~\cite{markus,pisarski} and has been compared~\cite{lacy,joe,koch} 
with commonly known fluids. Moreover, for fluids $\eta$ is either a 
nondiverging or a very weakly diverging quantity~\cite{rg,schafer} with 
an exponent $\sim -0.04$.
Thus, the fluidity of a hot viscous QCD matter remains an open as well as 
interesting problem that requires a much desired exploration of the fluid 
mechanics.

One can now ask what is the physical quantity that would reflect a good 
measure of fluidity of a hot viscous fluid or in this context how much relevant 
is the $\eta/s$, which is formed by two quantities, completely different 
in nature as $\eta$ is a dynamical transport coefficient whereas $s$ is a
thermodynamic one. A possible explanation lies within the general structure 
of any diffusion constant may it be thermal diffusion, ${\cal D}_{\kappa}$ 
related to the diffusive decay of longitudinal components of momentum 
fluctuation or the viscous diffusion, ${\cal D}_{\eta}$ related to the
diffusive decay of transverse components of momentum fluctuation of fluids. 
In this letter we try to address this important aspect of a hot fluid based on
fluid dynamics.

Dynamical properties of a many particle system can be investigated by
employing an external probe, which disturbs the system only slightly
in its equilibrium state, and by measuring the response of the system
to this external perturbation. A large number of experiments belong to
this category, such as studies of various line shapes, acoustic
attenuation, and transport behaviour. In all these experiments, one
probes the dynamical behaviour of the spontaneous fluctuations in the
equilibrium state. In general, the fluctuations are related to the 
correlation function, which provide important inputs for quantitative
calculations of complicated many-body system.

The density-density correlation function is defined as
\begin{equation}
\sigma_{nn}(\mathbf r,t) = \langle \delta n(\mathbf r,t)\delta 
n(\mathbf 0,0)\rangle \ \ ,
\label{corr}
\end{equation}
where the angular bracket denotes an equilibrium ensemble average, and
%\begin{equation}
$\delta n(\mathbf r, t)= n(\mathbf r, t) -\langle n \rangle ,$
%\label{devi}
%\end{equation}
is the local deviation of the number density $n(\mathbf r,t)$ from the 
equilibrium value of the number density. The spectral function 
can be obtained by Fourier transformation of (\ref{corr}) as
\begin{equation}
 \varrho_{nn}(\mathbf q,\omega)=\int d^3\mathbf r \int_{-\infty}^{\infty} dt 
\ \sigma_{nn}(\mathbf r,t) \ e^{-i(\mathbf q.\mathbf r - \omega t )} \ ,
\label{fourier}
\end{equation}
and the static correlation function can be defined as 
\begin{equation}
 {\tau}_{nn}(\mathbf q)=\int_{-\infty}^{+\infty}\frac{d\omega}{2\pi}
\varrho_{nn}(\mathbf q,\omega) \ . \label{static}
\end{equation}
Now, within hydrodynamical description the correlation function can 
be calculated only when wave number of the dynamical density fluctuation, 
${ q}$ is much smaller that the inverse correlation length, $\xi$
($q\xi<<1)$, or, equivalently wavelength, $\lambda$ of density fluctuation 
is appreciably larger than the correlation length ($\lambda>>\xi$). 
In the case of nonrelativistic
fluids, the dynamical density fluctuation based on Navier-Stokes 
equation~\cite{navier} in the hydrodynamic limit as well in the nonhydrodynamic 
limit ({\it viz.}, around critical point) has been studied~\cite{nonrel} for
Newtonian fluids in details. Recently, a relativistic generalisation of 
density fluctuation
is obtained~\cite{kunihiro} using dissipative relativistic fluid 
dynamics coupled with conserved quantities and
% the desired expression for 
the dynamical structure function is
\begin{eqnarray}
\varrho_{nn}(\mathbf q, \omega)/\tau_{nn}(\mathbf q)&=&
\left(1-\frac{1}{\gamma}\right) 
\frac{2{\cal D}_{\kappa}q^{2}}{\omega^{2}+({\cal D}_{\kappa}q^{2})^{2}} 
+\frac{1}{\gamma}
\nonumber \\ 
 \left\{\frac{{\cal D}_{s}q^{2}/2}{(\omega-c_{s}q)^{2}+
(\frac{{\cal D}_{s}q^{2}}{2})^{2}} \right.
&+&\left. \frac{{\cal D}_{s}q^{2}/2}{(\omega+c_{s}q)^{2}
+(\frac{{\cal D}_{s}q^{2}}{2})^{2}}\right \} ,
\label{struc}
\end{eqnarray}
where $\gamma$ is the ratio of specific heats, 
${\tilde C}_{P}=T(\partial S/\partial T)_P$ at constant 
pressure to 
${\tilde C}_{n}=T(\partial S/\partial T)_n$ at constant density
with $S$ is the equilibrium entropy per particle, $c_s$ is the speed of sound 
propagation with the damping constant ${\cal D}_s$ and 
${\cal D}_{\kappa}$ is the thermal diffusivity. 
The higher-order terms involving the 
quantities ${\cal D}_{\kappa}q/c_{s}$ 
and ${\cal D}_{s}q/c_{s}$ are neglected~\footnote{ 
A relativistic first order equation, such as that by Landau or by Eckart, 
is parabolic and formally violates the causality, and hence acausal. 
The causality problem can be circumvented with the inclusion of the 
second-order in derivative expansion~\cite{heinz,kunihiro,muronga}. However, 
this does not affect the spectral function of dynamical density fluctuation.}.
The thermal diffusivity is defined as,
\begin{equation}
 {\cal D}_{\kappa}= \frac{\kappa}{n_0 {\tilde C}_{P}} \ , 
\label{thermdiff}
\end{equation}
where $\kappa$ is the thermal conductivity and $n_0$ is the equilibrium
number density of fluid. Note that the nonrelativistic equivalence of
$n_0{\tilde C}_P$ in (\ref{thermdiff}) is $\rho_0 C_p$, where $\rho_0$ 
is the equilibrium mass density.

The sound wave  damping constant is given as
\begin{equation}
{\cal D}_{s}=
{\cal D}_{\kappa}(\gamma-1)
+\frac{\frac{4}{3}\eta+\zeta}{w_0}+
 c_s^2T^2\left (\frac{\kappa}{w_0}-2{\cal D}_{\kappa}
\alpha_P \right ) \ ,
\label{sounddiff}
\end{equation}
where $\eta$ and $\zeta$ are, respectively, shear and bulk viscosities,
$w_0$ is the equilibrium enthalpy density and $\alpha_P$ is the thermal
expansivity at constant pressure. The third term in (\ref{sounddiff})
is purely relativistic correction whereas the second term is the minimal 
relativistic one due to the appearance of $w_0$ instead of
$\rho_0$.

The dynamical structure function of density 
fluctuation for relativistic dissipative fluids in (\ref{struc})
is a sum of three Lorentzian line shapes with general form
$f(\omega)={2\Gamma}/[\Gamma^{2}+(\omega-\omega')^{2}]$,
centered at the frequency $\omega'$ with a half-width at half maximum 
given by $\Gamma$. The Rayleigh component, is centered about $\omega =0$, 
with a half width
\begin{equation}
 \Gamma_{R}={\cal D}_{\kappa}q^{2}\ \ . \label{rayleigh}
\end{equation}
The Brillouin doublets, are located symmetrically about $\omega=0$ at 
the frequencies $ \omega_{B}^{\pm}={\pm}c_{s}q$, 
each one with a half-width given by
\begin{equation}
\Gamma_{B}=\frac{1}{2}{\cal D}_{s}q^{2} \ \ . \label{brillouin}
\end{equation}
%ai-ks-hw1-hw2-ostm-comp.pdf
%anomaly-ostm.pdf
\begin{figure}%[t]
\includegraphics[height=3in,width=4in,clip]{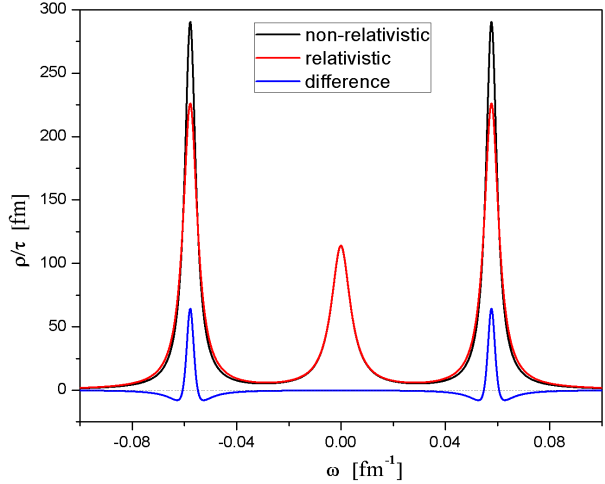}
\vspace*{-0.1in}
\caption{(Colour online) The structure function $\varrho_{nn}/\tau_{nn}$ as 
a function
of $\omega$ in the minimal relativistic fluid (red curve), nonrelativistic 
fluid (black curve) and their difference (blue curve) 
for parameters~\cite{kunihiro} $\ q=0.1 \ {\rm{fm}}^{-1}$, chemical potential
$\mu=200$ MeV, $T=200$ MeV, $\eta/(n_0S)=\zeta/(n_0S)=0.3$ and 
$\kappa T/(n_0S)=0.6$. The minimal relativistic correction shows up
in the Brillouin doublets  but not in the Rayleigh component centered 
at $\omega=0$ (see text). }
 \label{comparison}
\end{figure}

We note that in (\ref{struc}) the transverse component of the momentum
fluctuation decouples from the density fluctuation but admits a diffusive 
solution that relaxes exponentially with the viscous 
diffusion, ${\cal D}_{\eta}$ as  
\begin{equation}
{\cal D}_{\eta}=\frac{\eta}{w_0} \ \ , \label{viscdiff}
\end{equation}
which reduces to $\eta/\rho_0$ in the nonrelativistic limit.
The knowledge of the line widths in (\ref{rayleigh}) and (\ref{brillouin})
along with the three diffusion equations (\ref{thermdiff}),  
(\ref{sounddiff}) and (\ref{viscdiff}) are sufficient to determine
the three transport coefficients, $\kappa$, $\eta$ and $\zeta$. 

Now we note that the peak and the width of the Brillouin doublets obviously 
change due to the relativistic corrections of the sound mode, ${\cal D}_s$ 
in (\ref{sounddiff}). In contrast the Rayleigh peak as well as the width 
$\Gamma_R$ remain same for both nonrelativistic and relativistic fluids 
because the thermal diffusion ${\cal D}_{\kappa}$ does not change due to the 
equivalence between $n_0{\tilde C}_P$ and $\rho_0 C_P$. We demonstrate 
this aspect in Fig.~\ref{comparison} with only the minimal relativistic
corrections in (\ref{sounddiff}).  Further, the line width $\Gamma_R$ is 
controlled by the dominant behaviour of either $\kappa$
or ${\tilde C}_P$ in ${\cal D}_\kappa$. The
behaviour of ${\cal D}_s$ will control that of $\Gamma_B$ through the relative 
behaviours of $\kappa/{\tilde C}_n$, $\eta$ and $\zeta$. Also the strength 
of the various peaks in (\ref{struc}) depends on $\gamma$. 
This hydrodynamics predictions have been used in the limit
$T\rightarrow T_C$ to study the behaviour of relativistic fluid~\cite{kunihiro}
around critical point. The fundamental assumption of hydrodynamic, 
$q\xi<<1$, close to $T_C$ is no longer valid but it was argued that for a 
given $q$ there will be always a temperature range over which hydrodynamics
predictions should be reliable.  Around the critical point $\gamma$ diverges 
faster than the viscosities ({\em viz.}, $\zeta$), one can easily find from 
(\ref{struc}) that the sound mode gets attenuated and the only surviving 
soft mode is the diffusive thermal mode. This has important consequence as 
we will see below.

Generally, the long wavelength part of the density fluctuation  around
critical point is very intense that could induce a velocity gradient in 
the boundary of the fluid, which would dissipate energy. This dissipation 
can be interpreted in terms of the mode-mode coupling~\cite{mode,kawasaki}.
% and the various transport coefficient can be calculated. 
For instance, as discussed above, the heat flow mode (thermal diffusion) 
decays into shear viscous mode plus a thermal mode. One can obtain this decay 
mode in terms of the relevant transport coefficients. 

\begin{figure}%[t]
\includegraphics[width=4in,clip]{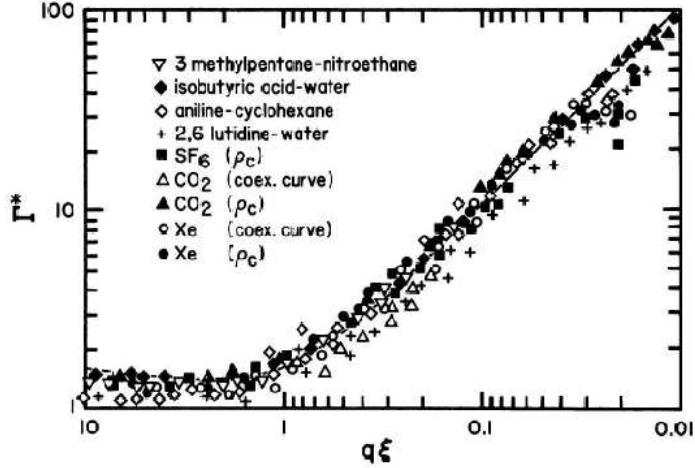}
%\vspace*{-0.3in}
\caption{$\Gamma^{*}$ as a function of $q\xi$ for several several  
substances that cover a wide range of molar mass, chemical structure and 
complexity and it is adopted from \cite{swinney}. Note that the measured 
data are from various experiments and the detailed references of those
can also  be found in \cite{data,swinney,NIST}. }
\label{compua1a2ostm}
\end{figure}

Information on critical behaviour of fluids are generally extracted from 
scattering experiments performed on the fluids~\cite{data}. To analyse these
data Swinney and Henry~\cite{swinney}, based on mode-mode-coupling theory, 
obtained a particular dimensionless combination of measured quantities 
for wide range of fluids obeying liquid-gas phase transition as
\begin{eqnarray}
\Gamma^{*}=\frac{6\pi\eta\Gamma_R}{Tq^3} \, . \label{scale}
\end{eqnarray}
Using the measured values of $\Gamma_R$, $\eta$ and the correlation length,
$\xi$ for 
different temperatures and scattering angles ($q=4\pi\sin(\theta/2)/\lambda$,
$\theta$ is the angle of scattering), obtained for various sample 
of multiple component fluids~\cite{data}, when plotted $\Gamma^*$ 
as a function of $q\xi$ all fluids obeying liquid-gas phase transition 
fall on a single universal curve as shown in 
Fig.~\ref{compua1a2ostm}. The single curve describes the critical behaviour
not only along the critical isochore and the coexistence curve, but also along
any thermodynamic path in the critical region. 
It is clear from Fig.~\ref{compua1a2ostm} that,
\begin{eqnarray}
\Gamma^{*}=\left \{ \begin{array}{ll}
                  \frac{1}{q\xi} & \mbox{ for $q\xi < 1$ \ 
                   (hydrodynamic domain),}  \label{domain} \\
                  1 & \mbox{ for $q\xi \geq 1$ \ \ \ (near critical point),}
                   \\
                  \end{array}
                 \right.
\label{gamma}
\end{eqnarray}
provided $\frac{6\pi \eta \Gamma_R \xi}{Tq^2}=1$, and using (\ref{rayleigh}) 
it becomes
\begin{equation}
{\cal F}_{\eta \kappa } 
=\frac{\eta}{({\cal D}_{\kappa}\xi)^{-1}\ T}=\frac{1}{6\pi}, \label{new}
\end{equation}
which is independent of the nature of the fluid despite the various 
microscopic details.
This is also consistent with Stokes formula for the mobility of a sphere
of radius $\xi$ moving through a viscous liquid with  a nondiverging 
shear viscosity, $\eta$. 

Now, the equality, $\Gamma^*=1$ in (\ref{domain}) basically holds
at and near the critical point, indicating the length scale matching 
around the critical point. This implies that instead of diverging without 
any upper bound, the correlation length of fluid becomes of the order of 
the wavelength of the density fluctuation ($\xi \sim \lambda$). On the
other hand this also suggests that ${\cal D}_\kappa\xi$ in together is
a nondiverging quantity around critical point since both of them are
expected to have same critical exponent~\cite{navier} but opposite in sign. 
The dimensionless quantity, ${\cal F}_{\eta\kappa}$ associated with the
viscous and the thermal modes together, could then be regarded as a good 
fluidity measure of a hot viscous fluid obeying liquid-gas phase transition, 
which remains at a universal value $1/(6\pi)$ for both nonhydrodynamical as 
well as in hydrodynamical region. 

We now note that the important quantity that enters in (\ref{gamma}) and 
(\ref{new}) is the Rayleigh width, $\Gamma_R$. The analysis, based on the 
Navier-Stokes theory presented in the beginning of this letter, clearly 
exhibits that the relativistic correction does not show up in the Rayleigh 
width $\Gamma_R$ in (\ref{rayleigh}). It indicates that the fluidity, 
${\cal F}_{\eta\kappa}$ is, obviously, independent of the choice of the 
({\em viz.}, relativistic or nonrelativistic) fluid dynamics. 
{\em This possibly indicates that a good fluid, either relativistic or 
nonrelativistic, is always a good fluid irrespective of the other 
details of the microscopic degrees of freedom.}  

In a heavy-ion collisions, the produced matter has a strong
flow initially along the beam axis. If the matter is viscous the viscosity 
then counteracts this by reducing the effective longitudinal pressure and 
thus enhancing it in the transverse plane. Therefore, the viscosity leads to 
a larger radial flow than that of the ideal fluid and also to an angular 
modulated radial flow ($v_2$) due to the velocity gradient.
This is indeed the situation in the experiments at RHIC~\cite{white,ellip}
that probably probed a QGP at a temperature little above $T_C$ with a small 
value of $\eta/s$~\cite{heinz,asis,lacy,joe} implying that the QCD matter  
produced in such experiments is a nearly perfect fluid and likely to obey
a liquid-gas phase transition with a critical end point~\cite{cep}. 
The spectral analysis for relativistic fluid in (\ref{struc}) or in 
Fig.~\ref{comparison} represents that a small density perturbation propagates
at the speed of sound and is eventually damped near the critical point. The
only surviving mode is the diffusive heat flow mode. The observed fluidity 
could be interpreted on the basis of mode-mode coupling 
theory as the decay of a longitudinal heat flow mode in QGP to a viscous 
and a thermal mode in which the transverse viscous mode reduces the thermal 
flow in the QGP. The QGP fluid produced in RHIC belongs to 
the same universality class defined by the fluidity measure in (\ref{new})
that interrelates the relevant transport coefficients, {\it viz.}, the 
shear viscosity $\eta$, and the thermal conductivity $\kappa$.  

We now recall that for fluids $\eta$ is either be nondiverging or weakly 
diverging
(with an exponent $\sim -0.04$)~\cite{schafer,rg}. One can expect
that $\eta/s$ should also be a nondiverging or weakly diverging quantity 
as entropy generally does not show any critical exponent. 
Surprisingly, for simple fluids as well as for QGP $\eta/s$ (as obtained 
from lattice QCD simulation~\cite{lattice} and 
also from the viscous hydrodynamic calculations~\cite{heinz,asis,lacy} 
in view of recent RHIC data~\cite{ellip}) shows an exponent 
(see Fig.3 of Ref.~\cite{lacy}) of the order of 'one'. 
%This is in contradiction with the general notion. 
This observed feature of $\eta/s$ can clearly be seen from (\ref{new}) as
\begin{equation}
\frac{\eta}{s}= {\cal F}_{\eta\kappa} \frac{\mu}{S}(\kappa\xi)^{-1},
\label{etas}
\end{equation}
where inverse of $\kappa\xi$ in together has an exponent $\sim 1.2$ for 
liquid-gas phase transition~\cite{navier}. 

For various fluids as discussed in Refs.~\cite{lacy,NIST}
one can parametrise $\eta/s$ with the reduced temperature, 
$\epsilon=|T/T_C-1|$, in the domain $T_C\leq T\leq 2T_C$ as 
\begin{equation}
\frac{\eta}{s} \sim (a\epsilon^\delta + b) \ ,
\end{equation}
where $a$ and $b$ are different for different fluids. 
For QGP~\cite{lattice,lacy} the elliptic flow data in RHIC~\cite{ellip}
restrict those as $\delta \sim 1$, $a\sim 0.64 \sim 2/3$ and 
$b\sim 0.1\sim 1/(4\pi)$, which describe the viscous mode despite
the various microscopic features in it. Obviously, 
the new fluidity measure ${\cal F}_{\eta\kappa}$ in (\ref{new}) then 
puts a constraint on the heat flow mode in QGP, which in turn restricts 
the associated transport coefficient $\kappa$ as 
\begin{equation}
\kappa = {\cal F}_{\eta\kappa}(a\epsilon^\delta+b)^{-1}\frac{\mu}{S\xi} \ ,
\end{equation}
where the correlation length, $\xi$ is related to various moments of
the net baryon number fluctuations~\cite{stephanov}.

In view of nearly perfect fluidity nature of QGP observed in RHIC we suggest 
through a new exponent-free fluidity that the thermal conductivity,
which is a measure of transport of energy by the particles, also becomes 
an important quantity to look at. This has not been discussed before in 
the literature in this context. Interestingly, it has been proposed to 
measure the thermal conductivity in RHIC with its 
upgradation~\cite{barbara}, which could then test the above estimation of 
this quantity both in hydrodynamical and nonhydrodynamical 
regions, and in turn our proposal of new fluidity measure. It will 
also be very interesting, if one can find a possible way to check this 
universal behaviour of fluidity through lattice QCD calculations and in 
nonperturbative models. 
The Large Hadron Collider (LHC) at CERN will produce a QGP at much higher 
temperature (at least twice of RHIC) and $\mu$ close to zero, which would 
then cool and pass through the critical region and it would be interesting 
to see whether there could still be a Rayleigh peak in this limit. 

We note that our analysis is based on the mode-mode coupling theory and
the relativistic Navier-Stokes theory. A basic assumption of this 
mode-mode-coupling formulation is the nondivergence of shear viscosity whereas 
a proper RG calculation reveals that shear viscosity has a very 
weak divergence governed by the exponent $-0.04$. On the other hand 
a Navier-Stokes theory is a kind of mean field equation and may not be very
appropriate to discuss the dynamics around critical point. The dynamics of 
the fluid in and around the critical point should, however, be described by 
the evolution equation derived from a microscopic theory. Nonetheless,
the present understanding of the critical dynamics of the produced
QCD matter in RHIC experiments is meagre and one needs to depend on a kind 
of mean field description to qualitatively understand the behaviour of 
the system in and around the critical point because the hydrodynamic 
variables become comparable to the time scale of an order parameter around 
the critical point due to critical slowing down.

\paragraph*{Acknowledgments:}
Authors are thankful to S. Chattopadhyaya for very fruitful discussion.
In particlar, RA is also thankful to A. Basu and A. Midya for very useful
discussion and help during the course of this work.

%%%%%%%%%%%%%%%%%%%%%%%%%%%%%%%%%%%%%%%%%%%%%%%%%%%%%%%%%%%%%%%%%%%%%%%%   

%%%%%%%%%%%%%%%%%%%%%%%%%%%%%%%%%%%%%%%%%%%%%%%%%%%

\end{document}